\documentstyle[aps,tighten,epsf,floats]{revtex}
\newcommand{\ben}{\begin{displaymath}}
\newcommand{\een}{\end{displaymath}}
\newcommand{\be}{\begin{equation}}
\newcommand{\ee}{\end{equation}}
\newcommand{\bea}{\begin{eqnarray}}
\newcommand{\eea}{\end{eqnarray}}

\begin{document}
% \draft command makes pacs numbers print
%\draft

\begin{center}
{\large \bf Renormalization of ${ }^1S_0$ NN scattering amplitude in Effective
Field Theory}
\end{center}

% repeat the \author\address pair as needed

\begin{center}
{ J. Gegelia}
\end{center}

\begin{center}
{\it INFN - Sezione di Ferrara, Via Paradiso 12, 
44100 Ferrara (FE), Italy \\
and \\
High Energy Physics Institute of TSU, University Str.. 9, Tbilisi 380086, 
Georgia}
\end{center}

\begin{center}

{ G. Japaridze}
\end{center}

\begin{center}
{\it Center for Theoretical Studies of Physical Systems, Clark Atlanta 
University, \\
Atlanta, GA 30314, USA}
\end{center}

\date{\today}
%\maketitle 

\begin{abstract}
Cutoff regularized subleading order ${ }^1S_0$ NN potential of  effective field
theory(EFT) is iterated using Lippmann-Schwinger equation. It is shown that 
the scattering amplitudes
calculated in cutoff and subtractively renormalized EFT are equal up to the 
accuracy of performed calculations.  
Non-perturbative renormalization, where part of divergences are absorbed into 
two contact interaction coupling constants with subsequent removal of 
regularization is also performed. Cutoff and dimensional regularizations 
both lead to finite but different results within this scheme.    
\end{abstract}
% insert suggested PACS numbers in braces on next line

{\it PACS numbers}: 03.65.Nk, 11.10.Gh, 12.39.Fe, 13.75.Cs. 

%\pacs{

\medskip
\medskip
\medskip

There has been significant development in effective field theory (EFT) 
approach to problems of multi nucleon systems originated by Weinberg's  
papers \cite{Weinberg:1990rz}, \cite{Weinberg:1991um}.  
(for recent developments see refs. \cite{Beane:2000fx},\cite{Bedaque:2000kn}). 
In Weinberg's approach power counting is formulated for effective 
potential and the scattering amplitude is calculated by substituting this 
effective  
potential into Lippman-Schwinger (or Schr\" odinger) equation.
In ref. \cite{Kaplan:1996xu} it has been argued that Weinberg's approach 
suffers 
from formal inconsistences. While this conclusion has been questioned 
\cite{Park:1999cu}, \cite{Gegelia:1998ee} a new explicitly consistent power 
counting (KSW power counting) has been suggested in ref. \cite{Kaplan:1998tg}.
In this approach pions are included perturbatively and it allows to perform  
all calculations   analytically which can be considered as a  considerable 
advantage. Unfortunately it has been realized that perturbation theory series 
do not converge in this approach \cite{Gegelia:1998ee},\cite{Cohen:1999jr},
\cite{Fleming:2000ee}. An attempt to combine 
the advantages of Weinberg's and KSW schemes has been made recently in 
ref. \cite{Beane:2001bc}. In that paper it has been confirmed that Weinberg's
power counting is formally inconsistent in the ${}^1S_0$ channel NN 
scattering problem. In particular, divergences which appear in leading order, 
can not be absorbed by leading order counter-terms. Different approach to the 
problem of non-perturbative renormalization of Lippmann-Schwinger equation 
has been suggested recently in ref. \cite{Eiras:2001hu}.     

In present letter,  which is the continuation of the work started in refs.
\cite{Gegelia:1999iu},\cite{Gegelia:1999ja}, we follow Weinberg's
approach and iterate sub-leading order ${ }^1S_0$ NN potential in the 
framework of Lippman-Schwinger equation. We note that the above mentioned 
inconsistency of Weinberg's power counting manifests itself only when it is 
applied to 
unrenormalized diagrams and counter-terms separately. We advocate the point of
view that the power counting should be applied to renormalized diagrams 
\cite{Park:1999cu},\cite{Gegelia:1998ee},\cite{Gegelia:1998gn},
\cite{Lepage:1999kt}. 
Let us emphasize that we do not attempt to resolve the above mentioned formal 
inconsistency of Weinberg's power counting, we believe that it can not be 
resolved within this scheme. We share the point of view that this problem
is irrelevant: if one applies the power counting to renormalized diagrams then 
there are no formal inconsistencies, renormalized diagrams are of the order 
which has been assigned by power counting and the approach is successful in 
studies of processes in few nucleon systems 
\cite{vanKolck:1999mw},\cite{Epelbaum:2000dj},\cite{Epelbaum:2001mx},
\cite{Epelbaum:2001fm}.
This scheme can be implemented using BPHZ renormalization procedure.
%Within this renormalization scheme power counting can be consistently applied 
%to the Lagrangian of the EFT (to decide which terms should be kept within 
%given order of calculations) as well as to Feynman diagrams. 
In BPHZ approach no counter-terms are included into the Lagrangian but 
instead the Feynman rules are accompanied by prescriptions for subtracting 
loop diagrams \cite{Collins:1984xc}. The subtracted diagrams depend on 
renormalization point(s). This dependence is cancelled by the renormalization 
point dependence of renormalised coupling constants so that the physical 
quantities are independent of the choice of subtraction point(s). The above 
mentioned problem of inconsistency that higher order counter terms contribute 
into the renormalization of low order diagrams translates now into the 
following problem: low order diagrams give large contributions into beta 
functions of coupling constants of higher order (counter) terms.       
There have been concerns that even if for some value of the renormalization 
point(s) the higher order coupling constants are tuned to be small, the slight 
change of this renormalization point(s) would result in large values for 
coupling constants and therefore power counting would  be badly violated. 
Note that these concerns are
based on perturbative estimations of beta-functions in the area where 
perturbation theory is not actually applicable. Hence there is no reason to 
expact that the assumption of naturalness which is an ingredient of Weinberg's
power counting is violated. To make any reliable 
conclusions about renormalization group behaviour of renormalized coupling 
constants non-perturbative analysis of this behaviour are required. We are not 
aware of any such analysis of
renormalized coupling constants in EFT with explicitly included pions. On the 
other hand the considerable success of cutoff EFT, which is based on 
Weinberg's approach 
\cite{vanKolck:1999mw,Epelbaum:2000dj,Epelbaum:2001mx,Epelbaum:2001fm}
suggests that renormalized coupling constants should be well-behaved.

Below we use Weinberg's power counting with BPHZ 
renormalization procedure and compare the results of subtractively 
renormalized and cutoff EFT for ${ }^1S_0$
NN scattering amplitude. We find that the results of two schemes coincide 
up to the order of the accuracy of given calculations. We also explore the 
possibility of 
non-perturbative renormalization via absorbing  part of divergences into the 
two available ``bare'' coupling constants with successive removal of 
regularization. 

In this letter we quote the  results, details of calculations will be 
presented elsewhere. 

We iterate subleading order potential substituting it into Lippman-Schwinger 
equation. To perform subleading order analysis we include two-derivative 
contact 
interaction part of the potential for ${ }^1S_0$ $NN$-scattering as it is 
enhanced in comparison with the two-pion exchange part \cite{Kaplan:1996xu}
(note that to renormalize the amplitude obtained by iterating leading order 
potential no contact terms with 
derivatives are required \cite{Kaplan:1996xu}, \cite{Beane:2001bc}) 
and therefore the sub-leading order potential has the following form:
\be
V(\mbox{\boldmath $p'$},\mbox{\boldmath $p$} )=C+C_2\left( 
\mbox{\boldmath $p'$}^2+\mbox{\boldmath $p$}^2\right)+
V_\pi \left( \mbox{\boldmath $p'$},\mbox{\boldmath $p$} \right)
\label{potential}
\ee
where  $C$ and $C_{2}$ are (``bare'') coupling constants, and
\be
V_\pi \left( \mbox{\boldmath $p'$},\mbox{\boldmath $p$} \right) =
-\frac{4\pi\alpha_\pi }
{\left( \mbox{\boldmath $p'$}-\mbox{\boldmath $p$}\right)^2+m_\pi^2},
\ \ \ \ 
\alpha_\pi =\frac{g_A^2m_\pi^2}{8\pi f_\pi^2}.
\label{vpi}
\ee

We need to extract the $S$-wave part from the solution to the  
Lippman-Schwinger equation for the $NN$ scattering amplitude $T$:
\be
T \left( \mbox{\boldmath $p'$},\mbox{\boldmath $p$} \right)=
V \left( \mbox{\boldmath $p'$},\mbox{\boldmath $p$} \right)+
\frac {M}{(2\pi)^3}\int d^3\mbox{\boldmath $q$} \
V \left( \mbox{\boldmath $p'$},\mbox{\boldmath $q$} \right) \
G(\mbox{\boldmath $q$}) \
T \left( \mbox{\boldmath $q$},\mbox{\boldmath $p$} 
\right)\equiv V \left( \mbox{\boldmath $p'$},\mbox{\boldmath $p$} \right)+
V \left( \mbox{\boldmath $p'$},\mbox{\boldmath $q$} \right) 
\otimes G(\mbox{\boldmath $q$})\otimes 
T \left( \mbox{\boldmath $q$},\mbox{\boldmath $p$} 
\right),
\label{lschweq}
\ee 
where 
$$
G(\mbox{\boldmath $q$})=\frac{1}{\mbox{\boldmath $k$}^2-
\mbox{\boldmath $q$}^2+i\epsilon },
$$ 
integration over "$\mbox{\boldmath $q$}$ - variable" $\frac {M}{(2\pi)^3}\int 
d^3\mbox{\boldmath $q$}$ is 
denoted by $\otimes$, $M$ is the nucleon mass and
$\mbox{\boldmath $p$}$ is on-mass-shell value of three-momenta. For physical 
amplitude $\mbox{\boldmath
$p'$}=\mbox{\boldmath $p$}$.

When substituted into Lippman-Schwinger equation  (\ref{lschweq}), potential 
 $V \left( \mbox{\boldmath $p'$},\mbox{\boldmath$p$} \right)$  generates 
divergences. 
 To regularize  potential we
introduce factorized form-factor into the contact interaction part of 
$V \left( \mbox{\boldmath $p'$},\mbox{\boldmath $p$} \right)$
and write:

\be
V_{R}(\mbox{\boldmath $p'$},\mbox{\boldmath $p$} )=\left[ C+C_2\left( 
\mbox{\boldmath $p'$}^2+\mbox{\boldmath $p$}^2\right)\right] 
\chi (\Lambda ,\mbox{\boldmath $p'$})\chi (\Lambda ,\mbox{\boldmath $p$}) +
V_\pi \left( \mbox{\boldmath $p'$},\mbox{\boldmath $p$} \right),
\label{potentialreg}
\ee
where $\chi $ is a regularization function. 
For definiteness let us use
\be
\chi (\Lambda ,\mbox{\boldmath $p$})=\frac {\Lambda^4}{\left( \Lambda^2+
\mbox{\boldmath $p$}^2\right)^2}.
\label{ff}
\ee

Using $V_{R}(\mbox{\boldmath $p'$},\mbox{\boldmath $p$} )$  instead of 
$V(\mbox{\boldmath $p'$},\mbox{\boldmath $p$} )$, we 
rewrite the  Lippman-Schwinger equation (\ref{lschweq}) into the matrix form:
\begin{eqnarray}
\nonumber
\left(
\begin{array}{cc}
T\left( \mbox{\boldmath $p'$},\mbox{\boldmath $p$}\right), & 0 \\
0, & 0
\end{array}
\right)=\chi \left( \Lambda ,\mbox{\boldmath $p'$} \right)
\left(
\begin{array}{cc}
1, & p'^2 \\
0, & 0
\end{array}
\right) \ 
\left(
\begin{array}{cc}
C, & C_2 \\
C_2, & 0
\end{array}
\right) \
\left(
\begin{array}{cc}
1, & 0 \\
p^2, & 0
\end{array}
\right)\chi\left( \Lambda ,\mbox{\boldmath $p$}\right)+
\left(
\begin{array}{cc}
V_\pi\left( \mbox{\boldmath $p'$},\mbox{\boldmath $p$} \right), & 0 \\
0, & 0
\end{array}
\right)+                \\
\nonumber
\frac{M}{(2\pi )^3}\int d^3\mbox{\boldmath $q$}
\left\{ \chi \left( \Lambda ,\mbox{\boldmath $p'$} \right)
\left(
\begin{array}{cc}
1, & p'^2 \\
0, & 0
\end{array}
\right) \ 
\left(
\begin{array}{cc}
C, & C_2 \\
C_2, & 0
\end{array}
\right) \
\left(
\begin{array}{cc}
1, & 0 \\
q^2, & 0
\end{array}
\right)\chi\left( \Lambda ,\mbox{\boldmath $q$}\right)+ 
\left(
\begin{array}{cc}
V_\pi\left( \mbox{\boldmath $p'$},\mbox{\boldmath $q$} \right), & 0 \\
0, & 0
\end{array}
\right)\right\} \times   \\
\left(
\begin{array}{cc}
G(\mbox{\boldmath $q$}), & 0 \\
0, & 0
\end{array}
\right) \ 
\left(
\begin{array}{cc}
T\left( \mbox{\boldmath $q$},\mbox{\boldmath $p$}\right), & 0 \\
0, & 0
\end{array}
\right)
\label{mlschweq}
\end{eqnarray}

\noindent
We solve (\ref{mlschweq}) for the ${ }^1S_0$ amplitude and obtain:

\be
T_0(p)=T_\pi+\frac{\left[ C+2p^2C_2 +C_2^2G_3(p)
\right] V_1(p)^2-2C_2V_1(p)V_2(p)\left[ 1+C_2G_2(p)\right]+
C_2^2G_1(p)V_2^2(p)}{\left[ 1+C_2G_2(p)\right]^2-\left[ C+2p^2C_2 +C_2^2G_3(p)
\right] G_1(p)},
\label{invt}
\ee
where $T_\pi$ is ${ }^1S_0$ wave solution of Lippman- Schwinger equation (\ref{lschweq}) for
$V \left( \mbox{\boldmath $p'$},\mbox{\boldmath $p$} \right)=
V_\pi \left( \mbox{\boldmath $p'$},\mbox{\boldmath $p$} \right)$. Functions  
$G_i(p)$ are given by infinite series as follows:

\begin{eqnarray}
\nonumber
G_1(p)=\chi^2\left( \Lambda ,\mbox{\boldmath $q$}\right)\otimes G\left( \mbox{\boldmath $q$}\right)+ 
\chi \left( \Lambda ,\mbox{\boldmath $q$}_1\right)\otimes  G\left( \mbox{\boldmath $q$}_1\right)\otimes 
V_\pi\left( \mbox{\boldmath $q$}_1,\mbox{\boldmath $q$}_2 \right)\otimes  G\left( \mbox{\boldmath $q$}_2\right)\otimes 
\chi \left( \Lambda ,\mbox{\boldmath $q$}_2\right)+ \\
\chi \left( \Lambda ,\mbox{\boldmath $q$}_1\right)\otimes  G\left( \mbox{\boldmath $q$}_1\right)\otimes 
V_\pi\left( \mbox{\boldmath $q$}_1,\mbox{\boldmath $q$}_2 \right) \otimes  G\left( \mbox{\boldmath $q$}_2\right)\otimes 
V_\pi\left( \mbox{\boldmath $q$}_2,\mbox{\boldmath $q$}_3 \right)\otimes  G\left( \mbox{\boldmath $q$}_3\right)\otimes 
\chi \left( \Lambda ,\mbox{\boldmath $q$}_3\right)+ \cdots
\label{g1exp}
\end{eqnarray}

\begin{eqnarray}
\nonumber
G_2(p)= \chi^2\left( \Lambda ,\mbox{\boldmath $q$}\right)\otimes 1+
\chi \left( \Lambda ,\mbox{\boldmath $q$}_1\right)\otimes  V_\pi\left( \mbox{\boldmath $q$}_1,\mbox{\boldmath $q$}_2 \right)
\otimes  G\left( \mbox{\boldmath $q$}_2\right)\otimes 
\chi \left( \Lambda ,\mbox{\boldmath $q$}_2\right)+  \\ 
\chi \left( \Lambda ,\mbox{\boldmath $q$}_1\right)\otimes 
V_\pi\left( \mbox{\boldmath $q$}_1,
\mbox{\boldmath $q$}_2 \right)\otimes 
G\left( \mbox{\boldmath $q$}_2\right)V_\pi\left( \mbox{\boldmath $q$}_2,
\mbox{\boldmath $q$}_3 \right)\otimes 
G\left( \mbox{\boldmath $q$}_3\right)\otimes
\chi \left( \Lambda ,\mbox{\boldmath $q$}_3\right)+ \cdots
\label{g2exp}
\end{eqnarray}

\begin{eqnarray}
\nonumber
G_3(p)=
\chi^2\left( \Lambda ,\mbox{\boldmath $q$}\right)\otimes 
G^{-1}\left( \mbox{\boldmath $q$}\right)+ 
\chi \left( \Lambda ,\mbox{\boldmath $q$}_1\right)\otimes 
V_\pi\left( \mbox{\boldmath $q$}_1,
\mbox{\boldmath $q$}_2 \right)\otimes  
\chi \left( \Lambda ,\mbox{\boldmath $q$}_2\right)+  \\
\chi \left( \Lambda ,\mbox{\boldmath $q$}_1\right)\otimes 
V_\pi\left( \mbox{\boldmath $q$}_1,
\mbox{\boldmath $q$}_2 \right) \otimes 
G\left( \mbox{\boldmath $q$}_2\right)\otimes 
V_\pi\left( \mbox{\boldmath $q$}_2,
\mbox{\boldmath $q$}_3 \right)\otimes 
\chi \left( \Lambda ,\mbox{\boldmath $q$}_3\right)+ \cdots
\label{g3exp}
\end{eqnarray}

\noindent
and the functions $V_1(\mbox{\boldmath $p$})$, $V_2(\mbox{\boldmath $p$})$ 
satisfy integral equations:
\be
V_1(\mbox{\boldmath $p$})=\chi \left( \Lambda ,\mbox{\boldmath $p$}\right)+
V_1\left( \mbox{\boldmath $q$}\right)\otimes 
G\left( \mbox{\boldmath $q$}\right)\otimes V_\pi\left( \mbox{\boldmath $q$},
\mbox{\boldmath $p$} \right),
\label{v1eq}
\ee

%$$
%V_1(p)=\chi \left( \Lambda ,\mbox{\boldmath $p$}\right)+
%\frac{M}{(2\pi)^3}\int d^3\mbox{\boldmath $q$} \ 
%\chi\left( \Lambda ,\mbox{\boldmath $q$}\right)
%G\left( \mbox{\boldmath $q$}\right)V_\pi\left( \mbox{\boldmath $q$},
%\mbox{\boldmath $p$} \right) +
%$$
%\be
%\left[ \frac{M}{(2\pi)^3}\right]^2 \int d^3\mbox{\boldmath $q$}_1 
%d^3\mbox{\boldmath $q$}_2 \ 
%\chi \left( \Lambda ,\mbox{\boldmath $q$}_1\right)
%G\left( \mbox{\boldmath $q$}_1\right)V_\pi\left( \mbox{\boldmath $q$}_1,
%\mbox{\boldmath $q$}_2 \right) 
%G\left( \mbox{\boldmath $q$}_2\right)
%\chi \left( \Lambda ,\mbox{\boldmath $q$}_2\right)
%V_\pi\left( \mbox{\boldmath $q$}_2,
%\mbox{\boldmath $p$} \right)+ \cdots
%\label{v1exp}
%\ee

\be
V_2(\mbox{\boldmath $p$})=
\chi \left( \Lambda ,\mbox{\boldmath $q$}\right)\otimes 
V_\pi\left( \mbox{\boldmath $q$},
\mbox{\boldmath $p$} \right)+
V_2 \left( \mbox{\boldmath $q$}\right)\otimes 
G\left( \mbox{\boldmath $q$}\right)\otimes 
V_\pi\left( \mbox{\boldmath $q$},
\mbox{\boldmath $p$} \right).
\label{v2eq}
\ee

%$$
%V_2(p)=\frac{M}{(2\pi)^3}\int d^3\mbox{\boldmath $q$} \ 
%\chi \left( \Lambda ,\mbox{\boldmath $q$}\right)
%V_\pi\left( \mbox{\boldmath $q$},
%\mbox{\boldmath $p$} \right)+
%\left[ \frac{M}{(2\pi)^3}\right]^2 \int d^3\mbox{\boldmath $q$}_1 
%d^3\mbox{\boldmath $q$}_2 \ 
%\chi \left( \Lambda ,\mbox{\boldmath $q$}_1\right)
%V_\pi\left( \mbox{\boldmath $q$}_1,
%\mbox{\boldmath $q$}_2 \right) 
%G\left( \mbox{\boldmath $q$}_2\right)
%V_\pi\left( \mbox{\boldmath $q$}_2,
%\mbox{\boldmath $p$} \right)+
%$$
%\be
%\left[ \frac{M}{(2\pi)^3}\right]^3 \int d^3\mbox{\boldmath $q$}_1 
%d^3\mbox{\boldmath $q$}_2 d^3\mbox{\boldmath $q$}_3\ 

%\chi \left( \Lambda ,\mbox{\boldmath $q$}_1\right)
%V_\pi\left( \mbox{\boldmath $q$}_1,
%\mbox{\boldmath $q$}_2 \right) 
%G\left( \mbox{\boldmath $q$}_2\right)V_\pi\left( \mbox{\boldmath $q$}_2,
%\mbox{\boldmath $q$}_3 \right) 
%G\left( \mbox{\boldmath $q$}_3\right)
%V_\pi\left( \mbox{\boldmath $q$}_3,
%\mbox{\boldmath $p$} \right)+ \cdots
%\label{v2exp}
%\ee
Expression (\ref{invt}) for the amplitude remains valid in the framework of 
dimensional regularization 
provided that  integrals are regulated using this regularization and 
$\chi $-function is substituted by $1$.

To give physical meaning to Eq. (7) we need to get rid off divergences, which 
show up in $\Lambda\to\infty $ limit. 
Note that the amplitude can be made finite using 
non-perturbative renormalization, i.e. renormalizing two available ``bare''
parameters $C$ and $C_2$ and taking $\Lambda\to\infty $ limit in 
non-perturbative expression of $T_0(p)$.

Let us  describe this procedure . 
We  fix $C$ and $C_2$ from two  conditions of finiteness of 
the amplitude at $p^2=0$ and $p^2=-\mu^2$ ($p=i\mu$). As $T_\pi (p)$ does not
contain divergences, we require the finiteness of 
$a(p)\equiv 1/\left[ T_0(p)-T_\pi (p)\right] $:

\be
a(0)=\frac{\left[ 1+C_2G_2(0)\right]^2-\left[ C+C_2^2G_3(0)
\right] G_1(0)}{\left[ C+C_2^2G_3(0)
\right] V_1(0)^2-2C_2V_1(0)V_2(0)\left[ 1+C_2G_2(0)\right]
+C_2^2G_1(0)V_2^2(0)},
\label{rencon1}
\ee

\be
a(\mu )=
\frac{\left[ 1+C_2G_2(i\mu )\right]^2-\left[ C-2\mu^2C_2 
+C_2^2G_3(i\mu )
\right] G_1(i\mu )}{\left[ C-2\mu^2C_2 +C_2^2G_3(i\mu )
\right] V_1(i\mu )^2-2C_2V_1(i\mu )V_2(i\mu )\left[ 1+C_2G_2(i\mu )\right]
+C_2^2G_1(i\mu )V_2^2(i\mu )}
\label{rencon2}
\ee
In (\ref{rencon1}) - (\ref{rencon2})  $a(0)$ and $a(\mu^2 )$ are 
finite.

Analyzing expressions (\ref{g1exp})-(\ref{v2eq}) it is straightforward to 
obtain following relations: 

\begin{eqnarray}
\label{g1}
G_1(p)&=&a_0 \Lambda +a_1 \ln \Lambda +a_2(p)+0\left( 1/\Lambda \right) \\
\label{g2}
G_2(p)&=&b_0 \Lambda^3 +b_1 \Lambda^2 +b_2(p)\Lambda +b_3(p) \ln \Lambda 
+b_4(p) +0\left( 1/\Lambda \right) \\
\label{g3}
G_3(p)&=&c_0 \Lambda^5 +c_1 \Lambda^4 +c_2(p)\Lambda^3+c_3(p)\Lambda^2 
+c_4(p)\ln \Lambda +c_5(p) +0\left( 1/\Lambda \right) \\
\label{v1}
V_1(p)&=&V_1(p)+0\left( 1/\Lambda \right) \\
\label{v2}
V_2(p)&=&v \Lambda +d(p)+0\left( 1/\Lambda \right),
\end{eqnarray}

where $a_0, a_1, b_0, b_1, c_l, c_1, v$ are constants independent of 
$\Lambda $ and   
$a_2(p), b_2(p), b_3(p), c_2(p)-c_5(p), V_{1}(p), d(p)$ are some functions of
 $p$ also 
independent of  $\Lambda $.

Substituting expansion (\ref{g1})-(\ref{v2}) into expressions  for $a(0)$ and 
$a(\mu)$  we solve the system of two equations for $C$ and $C_2$.
%and 
%solving the system of two equations for $C$ and $C_2$ we observe that 
%for $\Lambda\to\infty $ the emerging quadratic equation for $C_2$ has 
%solutions only if ({\bf where do we use  further requirement 
%(20)? NOWHERE, tu ginda amovagdot!  What the hell is $e_{0}$ in (20)?  IT IS 
%$a_0$})
%\be
%D= \Lambda^5 \left\{ a_0^2\left[ c_2(i\mu )-c_2(0)+2b_0\mu^2\right]
%\left[ a_2(0)-a_2(i\mu )+a(0) V_1^2(0)-a(i\mu ) V_1^2(i\mu )\right]\right\}
%\geq 0.
%\label{discr}
%\ee 
Substituting these solutions for $C$ and $C_2$ into expression for the 
amplitude 
(\ref{invt}) and  taking into account expansions   (\ref{g1})-(\ref{v2}) we 
obtain 
in the limit $\Lambda\to\infty $  amplitude which is finite:
\be
T_0(p)=T_\pi (p)+\frac{1}{\alpha(p)},
\label{nprampl}
\ee
where 
\begin{eqnarray}
\nonumber
& & \alpha (p)=\biggl\{ \left( a(\mu )V_1^2(\mu )+a_2(\mu ) \right) 
\left( c_2(p)-c_2(0)-4b_0p^2 \right)-
\left( a(0)V_1^2(0)+a_2(0) \right)\ \times
\\ 
& &\left( c_2(p)-c_2(\mu) 
-4b_0p^2-2 b_0\mu^2\right)+
a_2(p)\left( c_2(0)-c_2(\mu )-2b_0\mu^2\right) \biggr\}
/ \left[ \left( c_2(\mu )-c_2(0)+2b_0\mu^2\right)V_1^2(p)\right]
\label{treninv}
\end{eqnarray}

When using the same renormalization condition and dimensional regularization 
we 
obtain  finite result which differs from (\ref{treninv}) in analogy with 
the case of pionless 
EFT \cite{Beane:1998pk}. This difference between two results is due to the 
fact that dimensional 
regularization discards all power-law divergences, i.e. dimensional 
regularization is in fact not only regularization in standard sense but it 
also subtracts power-law divergences.
 
As it was argued in \cite{Gegelia:1998xr}, the above described renormalization
scheme is 
not consistent with EFT. To carry out the consistent renormalization procedure
it is necessary to absorb {\it all} divergences into coupling 
constants or in other words, subtract all divergent integrals. To do so 
let us  start from the 
observation that $G_{2}$, $G_{3}$ and $V_{2}$ can be expressed as follows: 

\begin{eqnarray}
\nonumber
G_2(p)&=&\Delta_2 +\Delta_\pi G_1(p), \\
\nonumber
G_3(p)&=&\Delta_3 +\Delta_\pi G_2(p), \\
V_2(p)&=&\Delta_\pi \ V_1(p),
\label{grelations}
\end{eqnarray}
where 
\begin{equation}
\Delta_2= 
\chi^2\left( \Lambda ,\mbox{\boldmath $q$}\right)\otimes 1, \ \   
\Delta_3=
\chi^2\left( \Lambda ,\mbox{\boldmath $q$}\right)\otimes 
G^{-1}\left( \mbox{\boldmath $q$}\right), \ \ 
\Delta_\pi =
\chi \left( \Lambda ,\mbox{\boldmath $q$}\right)\otimes 
V_\pi\left( \mbox{\boldmath $q$},
\mbox{\boldmath $p$} \right).
\label{deltadefs}
\ee  
Relations  (\ref{grelations}) are correct up to terms which are 
vanishing after renormalization is performed and regularization is removed ,
i.e. after  sub-divergences and overall divergences are substracted. Note that
these relations
become exact in dimensional regularization provided that in 
(\ref{deltadefs}) $\chi $-factors are substituted by 1.

Using relations  (\ref{grelations})  we express amplitude (\ref{invt}) as
\be
T_0(p)=T_\pi (p)+\frac{V_1^2(p)\ \left[ C -\Delta_\pi \ C_2 \left( 2
+\Delta_2 C_2\right) + 2\ p^2\ C_2 +\Delta_3 C_2^2
\right]}
{1 +2\Delta_2 C_2+ \Delta_2^2 C_2^2 -G_1(p)\left[ C -\Delta_\pi \ C_2 \left( 2
+\Delta_2 C_2\right) + 2\ p^2\ C_2 +\Delta_3 C_2^2 \right]} 
\label{tsimplified}
\ee
To renormalize amplitude  (\ref{tsimplified}) following standard BPHZ 
renormalization scheme we subtract divergent parts from 
$G_1(p)$, $\Delta_2$, $\Delta_3$ and  $\Delta_\pi$ and substitute $C$ and 
$C_2$ by their renormalized values $C_R$ and $C_{2R}$ 
%({\bf ? Which values, in
%what scheme are we now, 
%where this renormalised constants came from?})
and obtain:

\be
T_0^R(p)=T_\pi (p)+\frac{V_1^2(p)\ \left[ C_R -\Delta_\pi^R \ C_{2R} \left( 2
+F_2 C_{2R}\right) + 2\ p^2\ C_{2R} +\left( F_{31}+F_{32} \ p^2\right) C_{2R}^2
\right]}
{1 +2F_2 C_{2R}+ F_2^2 C_{2R}^2 -G_1^R(p) \left[ C_R -\Delta_\pi^R \ C_{2R} 
\left( 2 +F_2 C_{2R}\right) + 2\ p^2\ C_{2R} +\left( F_{31}+F_{32} 
\ p^2\right) 
C_{2R}^2 \right]} 
\label{trenormalized}
\ee
Here $F_2$, $F_{31}$ and $F_{32}$ are (arbitrary) finite parts. Note that 
arbitrary finite parts are also present in $\Delta_\pi^R$ and $G_1^R(p)$, the 
particular choice for the finite parts corresponds to the particular choice of
 renormalization 
condition.  

Expression (\ref{trenormalized}) agrees with the one for the renormalized  
amplitude, obtained in the framework of dimensional regularization 
\cite{Kaplan:1996xu}.
Amplitude obtained in \cite{Kaplan:1996xu} can be reproduced from 
Eq. (\ref{trenormalized}) by
appropriate choice of finite parts of $G_1(p)^R$ and $\Delta_\pi^R$, and 
by the conditions $F_2=F_{31}=F_{32}=0$.
Hence, as it was expected, dimensional and cutoff regularizations lead to the 
same 
results provided that the same renormalization condition is used.

\medskip

\medskip
Let us now turn to the cut-off EFT.
In this approach we absorb part of divergences into the 
 ``bare'' coupling constants $C$ and $C_2$ but
instead of considering  $\Lambda\rightarrow \infty $ limit,  we 
choose the value of cutoff parameter $\Lambda $ in such a way that the 
amplitude of cut-off theory reproduces the above calculated subtractively 
renormalized amplitude up to sub-sub-leading order.       

%$$
%\frac{1 - G_1(p)\ C}{V_1^2(p)\ C}+ \frac{ 2\left( -p^2\ V_1(p)+ \ V_2(p)+
%G_2(p)\ V_1(p)\ C - G_1(p)\ V_2(p)\ C\right)\ C_2}{V_1^3(p)\ C^2} 
%$$

We determine $C$ and $C_2$ from two conditions, the first of which is to 
absorbs all terms in Eq. (\ref{tsimplified}) for $p=0$, which are divergent 
if the $\Lambda\rightarrow \infty$ limit is taken. This condition can be 
written as:
\begin{eqnarray}
\nonumber
\frac
{1 +2\Delta_2 C_2+ \Delta_2^2 C_2^2 -G_1(p)\left[ C -\Delta_\pi \ C_2 \left( 2
+\Delta_2 C_2\right) + 2\ p^2\ C_2 +\Delta_3 C_2^2 \right]}
{V_1^2(0)\ \left[ C -\Delta_\pi \ C_2 \left( 2
+\Delta_2 C_2\right) + 2\ p^2\ C_2 +\Delta_3 C_2^2
\right]}=  \\
\frac
{1 +2\Delta_{2R} C_{2R}+ \Delta_{2R}^2 C_{2R}^2 -G_1^R(p)\left[ C_R 
-\Delta_{\pi R} \ C_{2R} \left( 2
+\Delta_{2R} C_{2R}\right) +\Delta_{3R} C_{2R}^2 \right]}
{V_1^2(0)\ \left[ C_R -\Delta_{\pi R} \ C_{2R} \left( 2
+\Delta_{2R} C_{2R}\right) +\Delta_{3R} C_{2R}^2
\right]}
\label{rencond}
\end{eqnarray}
\noindent
Another condition can be obtained by expanding Eq. (\ref{tsimplified}) in 
powers 
of $C_2$ and requiring the finiteness of the first order 
term in $\Lambda\rightarrow \infty $ limit. Taking into account 
Eq. (\ref{rencond}), this requirement 
can be expressed as
\begin{equation}
%\nonumber
\frac{C_2}{C^2}= \frac{C_{2R}}{C_R^2}
\label{c2}
\end{equation}

Substituting the solutions to Eqs. (\ref{rencond})-(\ref{c2}) into the 
(unrenormalized) solution of Lippman-Schwinger equation given by 
Eq. (\ref{invt}) 
and expanding in powers of $C_{2R}$ we obtain:  

\begin{eqnarray}
\nonumber
T_0(p)=T_\pi (p)+\frac{V_1^2(p)\ C_R}{1 
+ C_R\ \left( \Delta_1 - G_1(p)\ \right)} 
+ \frac{2\ V_1(p)\ \left( p^2\ V_1(p)-\Delta_\pi^R \ V_1(p)
-C_R \ \Delta_2^R\ V_1(p) \right) C_{2R}}
{\left( 1 + C_R \left[ \Delta_1 - G_1(p)\right] \right)^2} \\
+ \frac{2\ V_1(p)\ \left( \Delta_\pi V_1(p) -V_2(p) \right) C_{2R}}
{\left( 1 + C_R \left[ \Delta_1 - G_1(p)\right] \right)^2}
+ \frac{2\ V_1(p)C_{2R} C_R \left\{ \Delta_2 V_1(p)
-G_2(p) V_1(p)+G_1(p) V_2(p) 
\right\}}{V_1(0)\ \left( 1+\Delta_1\ 
C_R \right)\ \left( 1 
+ C_R \left( \Delta_1 - G_1(p)\right)\ C_R\right)^2 }+\cdots ,
\label{cutofft}        
\end{eqnarray}
where $\Delta_1$ contains divergent part of $G_1$ as well as finite part,  
the latter depending on the choice of renormalization scheme.
Note that fourth and fifth terms  of Eq. (\ref{cutofft}) are 
suppressed by an additional factor of $1/\Lambda$ in comparison with  
third term. Higher order terms in $C_{2R}$ which are 
indicated by dots are suppressed by higher orders of small parameter 
(three-momentum $p$, pion mass or the renormalization scale which we choose of 
the order of $p$). As our potential includes only leading and subleading order
terms, 
calculated amplitude can be trusted up to this order, provided that formally 
higher order contributions are indeed suppressed. Note that these higher 
order terms also contain factors like $p^4 \Lambda^2 C_{2R}^2$.
To suppress this kind of contributions as well as terms with 
inverse powers
of $\Lambda $,  we need to choose cutoff $\Lambda $ of the 
order of 
large scale parameter $Q$ which governs the behavior of $C_R$, $C_{2R}$ 
\cite{Park:1999cu}, \cite{Gegelia:1999iu}. With this choice of cutoff parameter
one can easily see that two amplitudes given by Eq. (\ref{trenormalized}) and 
Eq. (\ref{cutofft}) coincide up to 
the order of accuracy of our calculations the difference being of higher 
order. 
 
\medskip
\medskip
\medskip

In this work we have iterated cutoff regulated ${ }^1S_0$ 
partial wave NN 
potential of subleading order EFT and wrote a closed expression for the 
scattering amplitude. We have 
performed non-perturbative renormalization via absorbing part of 
divergences into two available coupling constants and taking the removed 
regularization limit. Within this renormalization scheme cutoff and 
dimensional 
regularizations both lead to finite results but these results differ from each
other. This difference between two results is due to the 
fact that dimensional regularization discards all power-law divergences. We 
argue that this kind of renormalization is not consistent with EFT approach.

We advocate the point of view that Weinberg's power counting should be 
applied to renormalized diagrams 
\cite{Park:1999cu},\cite{Gegelia:1998ee},\cite{Gegelia:1998gn},
\cite{Lepage:1999kt}, i.e. one should draw all (an infinite 
number of them for NN-scattering) relevant diagrams, renormalize and sum them 
up. This summation is performed by solving Lippman-Schwinger equation (or 
Schr\" odinger equation). In fact we have solved the regularized equation and 
performed renormalization by subtracting divergences in the solution. For the 
case considered in this work it is not difficult to demonstrate that expanding
the renormalized solution to the Lippmann-Schwinger equation we reproduce 
separate renormalized Feynman diagrams.
%, i.e. solving the equation and 
%renormalizing the solution by subtracting divergencies corresponds to 
%first renormalizing diagrams and after summing them up. 
Of course in general  the solutions to 
Lippmann-Schwinger equation (or Schr\" odinger equation) can contain more 
contributions than just the sum of (an infinite number of) perturbative 
diagrams, but note that there is no power counting for such  
non-perturbative contributions.    

We have implemented the renormalization procedure 
using BPHZ procedure. Within this scheme no counter-terms are included in the
Lagrangian and therefore it is free of formal inconsistences that to remove 
divergences from diagrams of given order calculations one needs to include 
contributions of counter-terms which are of higher order. Let us emphasize 
that we have not attempted to solve this problem of formal inconsistences. We
argue that this issue is irrelevant and it can be avoided by applying power 
counting to renormalized diagrams. The usage of BPHZ renormalization proves to
be very useful in dealing with this problem.

The main result of our work reads as follows.
While the power counting is consistently formulated in subtractively 
renormalised EFT, in most cases we are not able to solve regularized equations
and therefore it 
is technically very difficult (or rather impossible) to implement subtractive 
renormalization scheme. In practice  most of practically important calculations
are performed using cutoff EFT. For the potential considered in this work 
we  have compared the  amplitude of cutoff EFT with  
subtractively renormalized amplitude. The cutoff amplitude is equal 
to subtractively renormalized amplitude up to (including) sub-leading order, 
i.e. the difference between these two is of higher order than the order of 
the accuracy of our calculations (note that subtractively 
renormalized subleading order calculations reproduce the phase shift analyses 
for ${ }^1S_0$ NN scattering reasonably well \cite{Gegelia:1999ja}). 

We consider the present work as a demonstration of a general feature of 
cutoff EFT that it
reproduces the results of subtractively renormalized theory up to the order of
the accuracy of given calculations. This suggests that the 
considerable success of Weinberg's approach implemented in cutoff EFT
\cite{vanKolck:1999mw},\cite{Epelbaum:2000dj},\cite{Epelbaum:2001mx},
\cite{Epelbaum:2001fm} is by far not accidental.

%{\bf ACKNOWLEDGEMENTS}

%\newpage 
%\begin{thebibliography}{99} 

\end{document}